# A Radio Signal Modulation Recognition Algorithm Based on Residual Networks and Attention Mechanisms

Ruisen Luo[1], Tao Hu[1], Zuodong Tang[1], Chen Wang[2], Xiaofeng Gong[1], and Haiyan Tu[1]

[1]College of Electrical Engineering, Sichuan University, 24 South Section 1, One Ring Road, Chengdu, China 610065

[2]Department of Computer Science, Rutgers University -- New Brunswick, Piscataway, New Jersey, USA 08854

**Abstract—** To solve the problem of inaccurate recognition of types of communication signal modulation, a RNN neural network recognition algorithm combining residual block network with attention mechanism is proposed. In this method, 10 kinds of communication signals with Gaussian white noise are generated from standard data sets, such as MASK, MPSK, MFSK, OFDM, 16QAM, AM and FM. Based on the original RNN neural network, residual block network is added to solve the problem of gradient disappearance caused by deep network layers. Attention mechanism is added to the network to accelerate the gradient descent. In the experiment, 16QAM, 2FSK and 4FSK are used as actual samples, IQ data frames of signals are used as input, and the RNN neural network combined with residual block network and attention mechanism is trained. The final recognition results show that the average recognition rate of real-time signals is over 93%. The network has high robustness and good use value.

## Ⅰ Introduction

In recent years, with the rapid growth of wireless communication technology, the importance of signal modulation classification is growing. How to accurately identify signal modulation mode under low signal-to-noise ratio (SNR) has become a difficult problem in wireless communication. In recent years, on modulation recognition of communication signals, the commonly used methods include time-frequency analysis-based method, feature parameter extraction-based method, high-order cumulant-based method, statistical pattern recognition theory-based method, decision tree theory-based method, etc[1-2]. When each method is used individually, it is greatly affected by environmental noises, and the classification effect is not satisfying. Therefore, the combination of multiple methods for signal pattern recognition is a promising way to solve the signal modulation recognition problem.

In the early stage, there were few references [1] [2] about the modulation mode of recognition signal by deep learning. Sparse auto encoders based on ambiguity function (SAE-AF) [3] proposed the method of generating image by using fuzzy function, and recognized seven digital modulation modes by combining automatic encoder. In 2016, GNU was used to generate a set of data sets containing 11 kinds of common modulation modes, including 8 kinds of digital modulation and 3 kinds of analog modulation. [5] The importance of data sets in various aspects is discussed, and a set of standard data sets and a number of machine learning methods are proposed as the baseline on this data set, which is about 75% at high signal-to-noise ratio. Reference [6] discusses the recognition effect of VGG ResNet on the data set generated by VGG ResNet with 24 modulation modes. It proves once again that deep learning has greater potential in the field of modulation recognition. Reference [7] uses the data set provided in reference [5], discusses the effect of ResNet network under various hyperparameters, and combines ResNet with LSTM to improve the recognition rate, which is about 83% at high signal-to-noise ratio. Reference [8] explores the application of spectrum in modulation recognition. By Fourier transform of signal, the image of 100*100*2 can be obtained. From the result of data set used in this paper, the spectrum can also get good results. Reference [9] proposes a pre-training method, which combines IQ data with 4-order cumulative statistics to obtain features, and uses CNN + LSTM to obtain nearly 90% recognition rate at high SNR. Reference [10] compares Bayesian optimization with genetic algorithm (CNN), (Inception), (Resnet) and (LSTM) in data set RML 2018.01A. Reference [11] proposes that higher recognition rate can be obtained by combining two data sets with two CNNs. Reference [12] constructs a spectrum map for recognition combined with deep learning network, and uses the Gaussian kernel function for filtering. From the experimental results of [12], the spectrum map has a certain reference value, but because the data set is not clear, it cannot give an accurate definition. Reference [13] posted ResNet + LSTM in arXiv with recognition rate close to 90% at high SNR. The experiment used densenet to compare. From the experimental results, the experimental results show that the effect is less than 90%.

h-order cumulant features and artificial neural network is proposed, which mainly recognizes {FSK, PSK, ASK, QAM} signals. In [14], a method for estimating channel and noise parameters is proposed. The mixed likelihood ratio test is used to classify signals. Reference [15] For MPSK signal, a method of signal recognition in real channel is proposed based on the Fourier transform graphic characteristics of the second and fourth power of the signal. In [16], a robust AMC method using convolutional neural networks (CNN) is proposed. Modulation recognition of multi-signal based on high-level convolutional neural network has been achieved. Then another intelligent modulation recognition algorithm based on Neural Networks with sparse filtering is presented in [17]. In a word, there are some shortcomings in current research, such as requiring high signal-to-noise ratio and high algorithm complexity.

In this paper, a signal pattern recognition algorithm based on RNN, which combines residual block network and attention mechanism, is proposed. Firstly, residual block network is used to solve the problem of gradient disappearance during network training. In training, the data can be concentrated in the region with the greatest gradient of activation function. Through the attention mechanism, the obvious part of signal change is highlighted, which achieves the goal of network optimization. Finally, under the optimization of these two modules, a well-performance RNN is trained and the radio modulation signal is well recognized. RadioML2016.10 radio signal standard data set is used as the research object. The data set covers 11 radio modulated signals with different modulation modes. Each signal has a data frame structure of 2*128, totaling 200,000. According to a certain proportion, the data in the data set are divided into training samples and test samples. The experimental results show that the average recognition rate of the network can reach more than 90% when the signal-to-noise ratio is greater than 5 dB, which surpasses most existing algorithms.

## Ⅱ Methods and Materials

### 2.1. Residual network

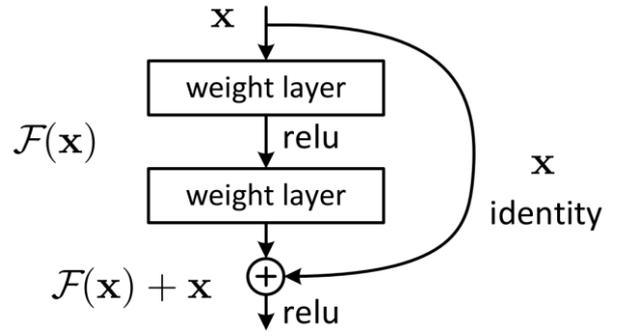

Fig.1 Residual block[]

Layers of networks that contain a short cut connection are called residual blocks. Residual networks become more easily optimized by adding short cut connections.

As shown in Figure 1, where x presents input, F(x) represents the output of the residual block before the second level activation function, that means $F(x) = W_2\sigma(W_1 x)$, where $W_2$ and $W_1$ represent weights of Layer 1 and Layer 2, $\sigma$ represents the ReLU activation function. Finally, the output of the residual block is $\sigma(F(x) + x)$.

Two residual blocks are connected in series. If there is no residual, it will be found that with the deepening of the network, the training error first decreases and then increases. In theory, the training error is smaller and better. As for residual network, with the increase of layers, the training error decreases. This way can reach the deeper layer of the network, help to solve the problem of gradient disappearance and gradient explosion, let us train deeper network while ensuring good performance.

### 2.2. Radio Signal Modulation Recognition Based on Attention mechanism LSTM Cyclic Neural Network

Radio signal pattern recognition model is usually built in the framework of Encoder-Decoder, that is, coding-decoding framework. This framework is mainly used to solve the problem of seq-2-seq, that is, when the input and output sequences are not equal.

The working principle of Encoder-Decoder framework is to encode input x, convert it into semantic code C through non-linear transformation, and decoder decodes the semantic code c to output the target sentence y. It can be seen that the Encoder-Decoder model generates the target sentence y by directly encoding and decoding the input sentence x, so the model pays equal attention to each input.

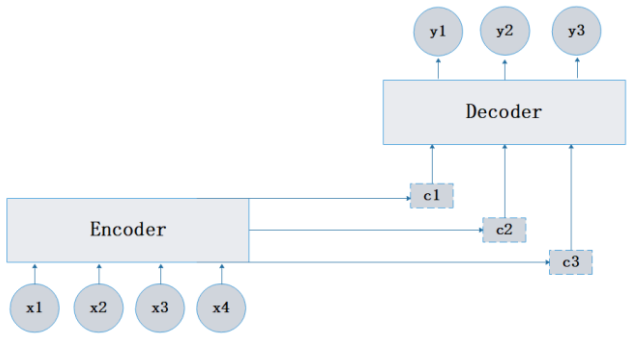

Fig.2 Attention Model Framework[]

The introduction of attention mechanism makes the model more attentive and can inject more attention into the noteworthy areas.

Compared with the CNN network mentioned, the cyclic neural network has certain memory function, but it often forget the information in the far ends. Therefore, LSTM units with three gate units are introduced on the basis of cyclic neural networks to filter and transmit information far away. Reduce the recognition error caused by the "forgetting" of the cyclic neural network.

When building a speech keyword recognition model using LSTM cyclic neural network, to use the convolution neural network introduced above to extract the features of the transformed two-dimensional image. Then the LSTM cyclic neural network is used to learn and map the output results. The basic flow chart is shown in Figure 2.

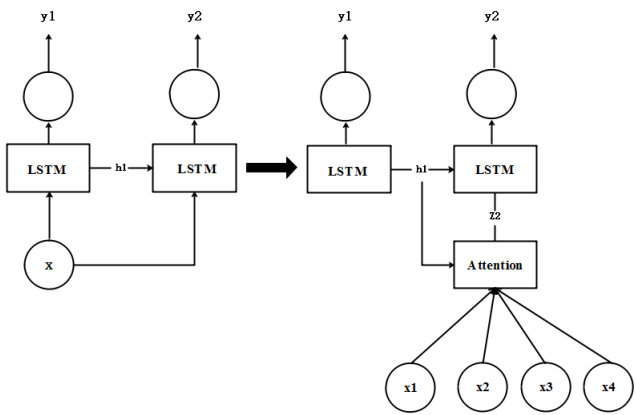

Fig.3 Basic Flow Chart of LSTM Cyclic Neural Network Model[]

In this paper, the bi-directional LSTM cyclic neural network is combined with the convolution neural network. At first, the Meyer spectrum is extracted by two convolution layers. After dimensionality reduction, the output of the last convolution layer is input into the bi-directional LSTM cyclic neural network. The number of LSTM units is 64. Finally, the output layer of LSTM is transferred to the full connection layer for classification.

## 2.3. Residual Networks and Attention Mechanisms in RNN

In order to solve the problem of gradient disappearance in the training process, the residual block network in the network plays a role of gradient concentration. The attention mechanism in the network is that the feature extraction of the network pays more attention to the difference of the characteristics of different types of signals, which enhances the robustness of the network. The proposed network structure is shown in Figure 4.

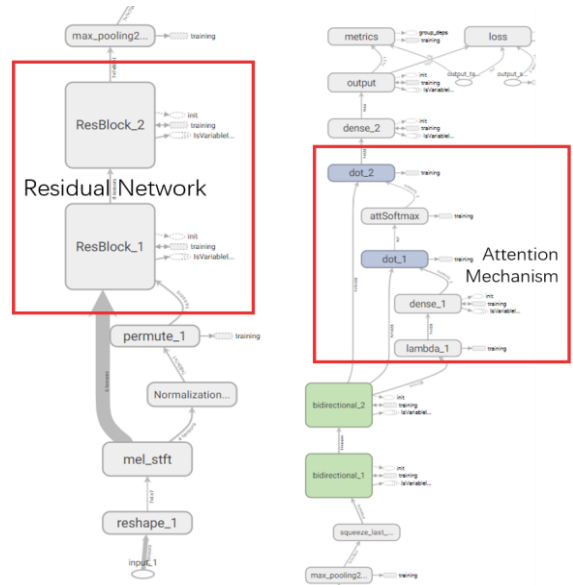

Fig.4 General block diagram of network structure

All the networks in this study introduce residual block networks and attention mechanism on the basis of traditional cyclic neural networks. The training data are 2 *8196 IQ data frames. Firstly, after shaping the data, the residual network composed of two residual blocks is input. After training the parameters of the residual blocks, the data are input into a convolution module, and the data are preliminarily reduced in dimension and extracted in feature. Then, the data are imported into a model of circular neural network with attention mechanism. It is modified on the basis of bidirectional LSTM cyclic neural network, that is, extracting the output vector of the last LSTM layer and using the dense layer projection as the query vector to identify which part of the signal is most relevant. Because the useful information in the signal file is in the middle, we choose to use the vector of LSTM output. This choice is arbitrary and any vector should work because the double-stacked LSTM layer should be able to carry enough memory. Finally, the weighted average of LSTM output is input to three full connection layers for classification.

### III Results and discussions

In order to verify the classification performance of various digital modulation signals using network and classification methods. Firstly, The database used in this study is RadioML2016.10 standard data set, which contains 11 signals of different modulation types. Then each data generates a data frame with the size of 2*128, which is imported into the built network for training. Then the test data are input into the trained network, and the network recognition results are compared and analyzed.

### 3.1. Signal IQ data acquisition

RadioML2016.10 standard radio signal data set is used for training and testing data. The data set includes 11 different types of modulation signal data. Each data frame is 2 *128 in size, totaling about 200,000 frames. A single data sample is used as the input of the network, and the training data set and the test data set are divided according to a certain proportion. IQ data diagrams of various signals are shown in Fig. 5.

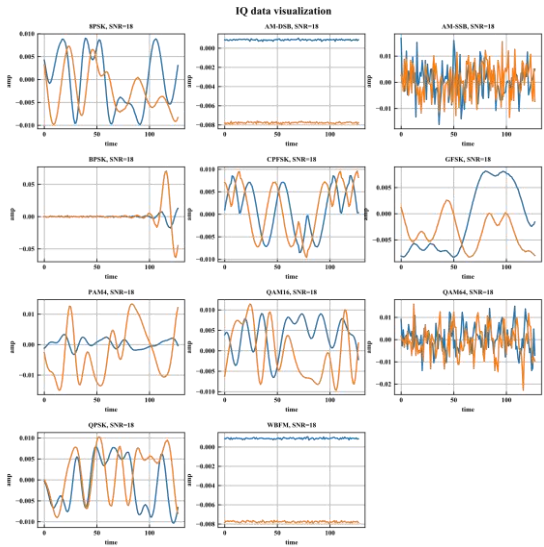

Fig.5 Standard Data Set IQ Data of Various Modulation Signals

The acquired data sets are sorted out and input, and the IQ data of each kind of modulation signal is marked. Then the data are input into the network, initialize the network parameters, and train them.

### 3.2. Network Recognition Results

By inputting all kinds of signals from the standard data set into the trained neural network, the confusion matrix diagrams of signal recognition accuracy under different signal-to-noise ratios are obtained. The vertical coordinate represents the real label of the test data and the abscissa represents the predicted result of the output data. The shade of each colour indicates the accuracy of recognition, and the darker the colour, the higher the accuracy. The results show the confusion matrix diagram of signal-to-noise ratio from -12dB to 10dB, as shown in Fig. 5.

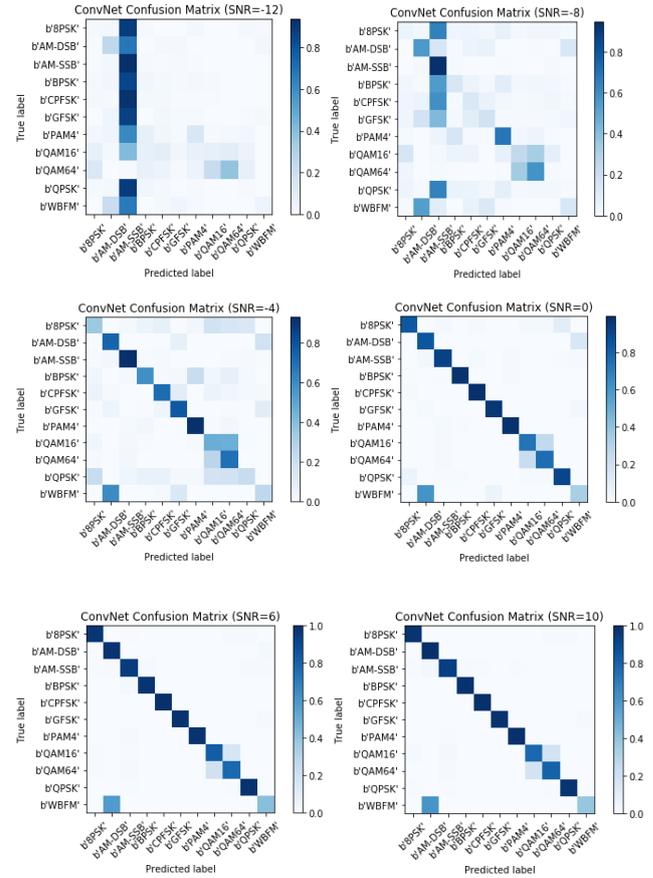

Fig.5 Diagram of Signal Recognition Accuracy Confusion Matrix

As can be seen from Figure 5, when SNR is - 12dB, all kinds of signals are basically recognized as AM - DSB signals; when SNR is - 8dB, the recognition accuracy of each signal diverges to different signals, which is still confused. When the signal-to-noise ratio is - 4dB, the recognition accuracy of each signal concentrates on the correct label, but the accuracy is generally not high. When the SNR is 0 dB, the classification of the predicted signal basically coincides with the label of the test sample, and the accuracy of signal recognition increases significantly; when the SNR is greater than 0 dB, the accuracy of signal recognition is more concentrated. Therefore, the accuracy of signal recognition increases with the increase of SNR. When the SNR exceeds 0 dB, the effect of signal recognition is good.

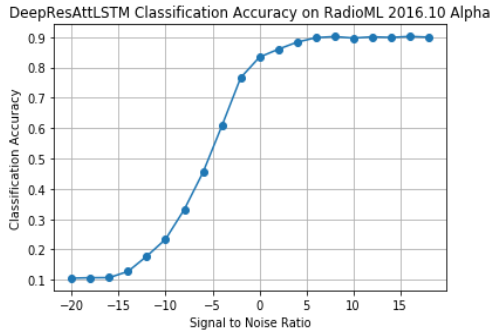

Fig.6 Statistical Chart of Signal Average Recognition Rate

The average recognition accuracy of all samples in the standard database is calculated. The average recognition rate chart of the network on -20dB to 20dB is shown in Figure 6. As can be seen from figure 6, the average recognition rate increases with the increase of SNR. After SNR is greater than 5dB, the average recognition rate can quickly reach more than 90%, which proves that the network can recognize signals accurately under the condition of low SNR.

## IV  Conclusion

In this paper, a signal pattern recognition algorithm based on RNN, which combines residual block network and attention mechanism, is proposed. Firstly, residual block network is used to solve the problem of gradient disappearance during network training. In training, the data can be concentrated in the region with the greatest gradient of activation function. Through the attention mechanism, the obvious part of signal change is highlighted, which achieves the goal of network optimization. Finally, under the optimization of these two modules, a well-performance RNN is trained and the radio modulation signal is well recognized. RadioML2016.10 radio signal standard data set is used as the research object. The data set covers 11 radio modulated signals with different modulation modes. Each signal has a data frame structure of 2*128, totaling 200,000. According to a certain proportion, the data in the data set are divided into training samples and test samples. The experimental results show that the average recognition rate of the network can reach more than 90% when the signal-to-

Preprint. Work in process.

noise ratio is greater than 5 dB, which surpasses most existing algorithms.